*Title:* **Development and Validation of the $^7$Li(p,n) Nuclear Data Library and Its Application in Monitoring of Intermediate Energy Neutrons**

*Author(s):* Alexander PROKOFIEV, Mark CHADWICK, Stepan MASHNIK, Nils OLSSON, and Laurie WATERS

# Development and Validation of the $^7$Li(p,n) Nuclear Data Library and Its Application in Monitoring of Intermediate Energy Neutrons

Alexander PROKOFIEV[1,2,*], Mark CHADWICK[3], Stepan MASHNIK[3], Nils OLSSON[4,5], and Laurie WATERS[3]

[1]*The Svedberg Laboratory, Uppsala University, Box 533, S-751 21 Uppsala, Sweden*
[2]*V.G. Khlopin Radium Institute, 2oi Murinskiy Prospect 28, Saint-Petersburg 194021, Russia*
[3]*Los Alamos National Laboratory, Los Alamos, New Mexico 87545*
[4]*Department of Neutron Research, Ångström Laboratory, Uppsala University, Box 525, S-751 20 Uppsala, Sweden*
[5]*Swedish Defense Research Agency (FOI), S-172 90 Stockholm, Sweden*

Systematics have been created for neutron spectra from the $^7$Li(p,n) reaction at 0° in the 50-200 MeV proton energy region. The available experimental data in the continuum part of the spectra show satisfactory overall agreement with a representation based on the phase-space distribution corresponding to the three-body breakup process $^7$Li(p,n $^3$He)α, with empirical correction factors, which depend regularly on incident energy. Validation of the systematics included folding of the predicted neutron spectra with standard $^{238}$U neutron fission cross section. Modeled in this way distributions of neutron-induced fission events agree reasonably with experimental data.

## I. Introduction

Despite the MeV energy region, a truly monoenergetic neutron source is not feasible at the intermediate energies. Among "quasi-monoenergetic" sources, the ones based on the $^7$Li(p,n) reaction[1-4] are widespread. Most of them utilize neutrons that propagate straight ahead, and therefore only spectra at 0° are further considered.

A number of high-energy neutron studies are intrinsically not capable of neutron energy selection. The examples are radionuclide production cross section measurements[5] and virtually all applied research, e.g., single-event effect studies[6,7] and calibration of neutron dosimeters[8,9]. Information on the neutron spectrum can be provided by simultaneous or dedicated beam characterization measurements[4] with time-of-flight (TOF) techniques. However, the TOF separation of the high-energy peak from the tail is demanding to the length of the neutron flight path, the time structure of the primary beam, and the time resolution of the experimental technique.

An alternative approach is to calculate the neutron spectrum, together with other reaction characteristics, using a nuclear reaction model code. It is implemented in a recent publication of Mashnik, *et al.*[10], that describes the first release of the $^7$Li(p,n) nuclear data file for incident proton energies $E_p \leq 150$ MeV, included in the LA150 library[11] and intended for the use in the MCNPX transport code. The evaluation[10] satisfactorily reproduces experimental data for $E_p < 40$-50 MeV, but there is a major disagreement at higher energies (see Sect. II).

In view of the difficulties associated with both mentioned approaches, semi-empirical systematics of the neutron spectra in the $E_p = 50$-200 MeV region have been developed in the present work (Sect. III). Validation of the systematics (Sect. IV) includes folding of the predicted neutron spectra with standard $^{238}$U neutron fission cross section[12,13]. Modeled in this way distributions of neutron-induced fission events are compared with experimental data.

## II. Review of the past work

Measurements of neutron spectra from the $^7$Li(p,n) reaction in the $E_p = 50$-200 MeV region were performed by a number of groups[1-4, 14-18] (see **Fig. 1**). The high-energy peak corresponds to the $^7$Li(p,n) reactions that leave the $^7$Be nucleus in the ground state or in the first excited state at 0.43 MeV. None of the considered experiments was able to resolve these two transitions, and, on the other hand, most applications need not such detailed information. The low-energy tail component may result from a few different physical mechanisms. The first and inevitable contribution is due to the properties of the nuclear reaction employed for neutron production. Then, interactions of primary and scattered protons, as well as of neutrons produced in the target, with the beam transport system, the walls, and other material in the surroundings, with subsequent propagation and slowing-down of resulting neutrons, may contribute. The latter components influence primarily the low-energy end of the spectrum and may cause its unwanted variation with experimental conditions, e.g., accelerator mode of operation[2,4]. On the other hand, neutron spectra measured by different groups and at different facilities exhibit an overall good agreement at the higher energy part, which is, therefore, likely to be dominated by fundamental properties

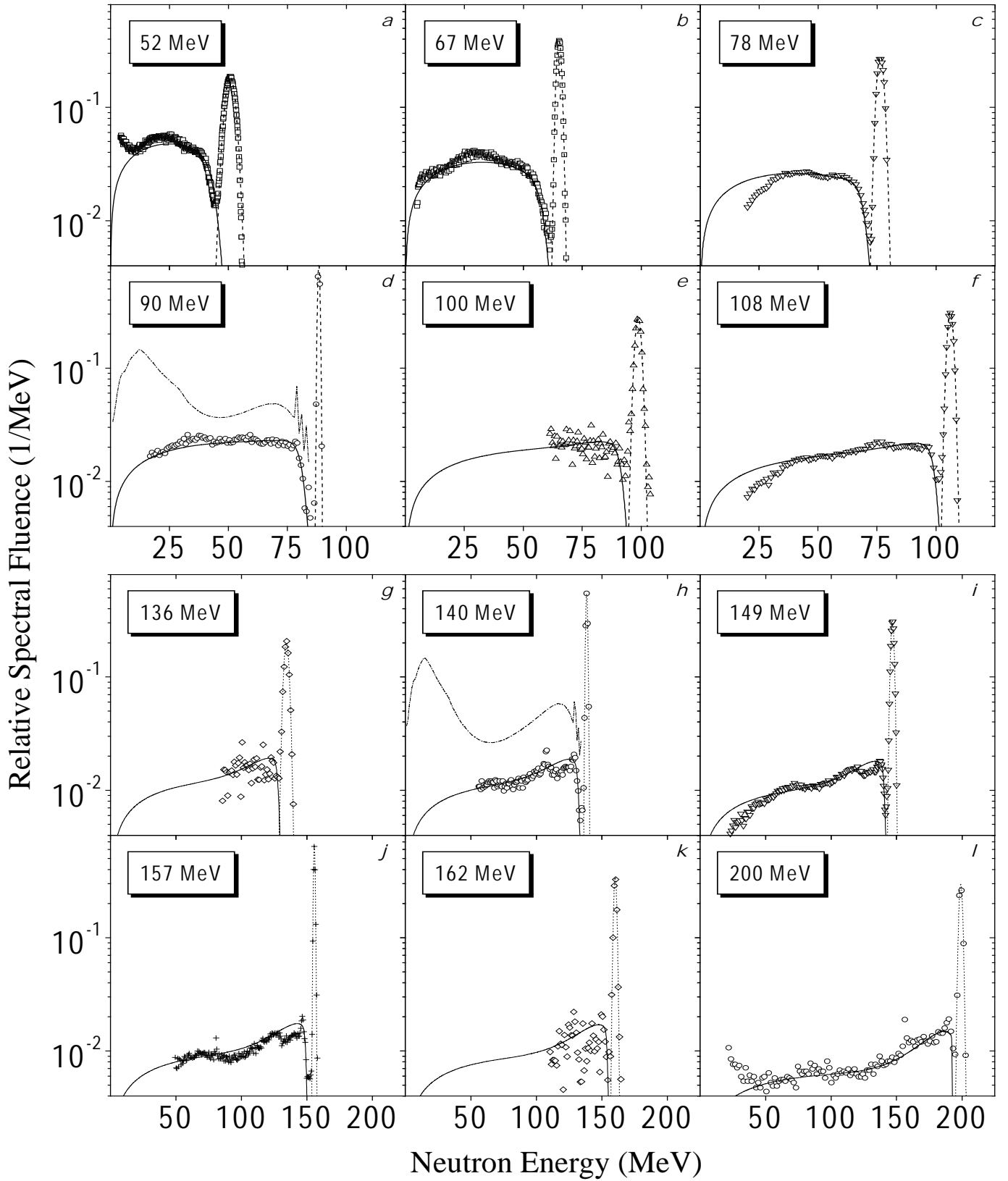

**Fig. 1.** Neutron spectra from the $^7$Li(p,n) reaction at 0°. The quoted energy corresponds to the average energy of the incident protons in the $^7$Li target. The symbols represent experimental data of Condé, *et al.*[1] (up triangles), Baba, *et al.*[2] (squares), Nakao, *et al.*[3] (down triangles), Byrd, *et al.*[15] (circles), Rönnqvist[16] (diamonds), and Scobel, *et al.*[17] (crosses). The dashed lines are Gaussian fits to the high-energy peaks. The solid lines show predictions of the systematics of the present work for continuum neutron production. The dash-dotted lines show predictions by the current LA150 library[10]. All spectra are normalized so that the area under the high-energy peak is unity.

of the $^7$Li(p,n) reaction.

Examples of predictions by the current LA150 library[10] for the continuum part of the spectrum are shown in Fig. 1d and 1h by dash-dotted lines. The observed disagreement is partly due to the underestimation of the high-energy peak cross section in the evaluation[10]. In addition, production of low-energy neutrons ($E_n \leq$ 40-50 MeV) is overpredicted by up to an order of magnitude, while the experimental spectra, e.g., [3] and [15], agree reasonably to each other. A possible reason of the difficulties is that the model calculations[10] based on statistical preequilibrium and equilibrium decay theories are inadequate for such a light nucleus as $^7$Li.

## III. The systematics for the neutron spectra

Baba, et al.[2] employed a phase-space distribution[19] corresponding to the three-body breakup process $^7$Li(p,n $^3$He)α for description of the neutron spectra obtained in the same work at $E_p$ = 40-90 MeV. Satisfactory overall agreement with the experimental data was observed, except of the highest and the lowest neutron energy region.

In the present work, an attempt was made to apply the similar description to all available experimental data in the $E_p$ = 50-200 MeV region. It has turned out that the middle part of the spectra can be successfully reproduced by the phase-space distribution[19]. The data at low neutron energies ($E_n$ < 10-20 MeV) are sparse, discrepant, and often influenced by various problems mentioned above, so the phase-space distribution still seems to be the best physically justified estimate. Finally, the high-energy part of the continuum spectrum (i.e. below the high-energy peak) deviates systematically from the phase-space distribution. The latter overpredicts the neutron production for $E_p$ < 70 MeV and underpredicts it for higher incident energies. Following an idea of Neumann[20], this effect has been taken into account by an empirical correction, as described below.

Assuming the area under the high-energy peak is unity, the continuum part of the neutron spectrum is represented as:

$$\varphi(E_n) = \varphi_{PS}(E_n) R(E_n), \quad (1)$$

where $\varphi_{PS}$ is the phase-space distribution[19] normalized so that the square under it is equal to unity, and $R(E_n)$ is an empirical factor. The latter takes into account the deviation of the high-energy part of spectrum from the phase-space distribution and also characterizes the abundance of the continuum part of the spectrum relative to the high-energy peak:

$$R(E_n) = R_0 \frac{1 + R_1 \exp\frac{E_n - E_0}{E_1}}{1 + \exp\frac{E_n - E_0}{E_2}}, \quad (2)$$

where $R_0, R_1, E_0, E_1,$ and $E_2$ are fitting parameters. They depend smoothly on $E_p$, and the following systematics have been suggested:

$$R_0 = q_1 + q_2 \exp(-E_p / q_3), \quad (3)$$

$$R_1 = \begin{cases} q_4(E_p - q_3), & \text{if } E_p > 68\,\text{MeV}, \\ 0, & \text{if } E_p < 68\,\text{MeV}, \end{cases} \quad (4)$$

$$E_0 = E_p - q_5, \quad (5)$$

$$E_1 = q_6 \{1 - \exp[-q_7(E_p - q_3)]\}, \quad (6)$$

$$E_2 = q_8 - q_9 E_p, \quad (7)$$

where $q_1$=0.8091, $q_2$=2.228, $q_3$=66.62 MeV, $q_4$=0.04521, $q_5$=8.097 MeV, $q_6$=19.07 MeV, $q_7$=0.0463 MeV$^{-1}$, $q_8$=2.638 MeV, and $q_9$=0.01250 are universal parameters obtained with the least-squares method from the experimental spectra [2, 3, 15, 17]. The other available data[1, 4, 14, 16, 18] were not used because of various reasons (insufficient statistics, energy resolution, and/or covered energy region; interference from surroundings). Predictions of the systematics are shown as solid lines in Fig. 1. They are in good overall agreement with experimental data.

In order to keep the systematics simple, no attempt was made to reproduce the fine structure in the high-energy part of the spectrum due to the excitation of low-lying discrete levels in $^7$Be nuclei. On the other hand, the spectral neutron fluence, integrated over the part of the spectrum where the structure is seen, is reproduced well by the systematics.

## IV. Validation of the systematics

The fraction of high-energy peak in the neutron spectrum has been calculated from the above-described systematics and plotted in **Fig. 2a** versus incident proton energy. In order to compare the predictions with experiment, the ratio of calculated and experimental data[2, 3, 15, 17] was calculated for each data set, and its average value was used for reconstruction of the missing part of the experimental spectrum. It is seen that the fraction of the peak neutrons amounts to about 40% and depends weakly on $E_p$. The root-mean-square difference between the calculated data and those deduced from the experiment is less than 7%.

The neutron spectra predicted by the systematics were folded with neutron-induced fission cross section of $^{238}$U, adopted as a neutron standard[12, 13]. Modeled in this way distributions of neutron-induced fission events were used for calculation of the fraction of fissions induced by high-energy peak neutrons. The results are compared with experimental data obtained at the TSL neutron facility[21] (see **Fig. 2b**). The root-mean-square deviation of the calculation from the experiment is about 15%. Taking into account uncertainties in the calculated and experimental data, their mutual agreement is considered as reasonable. Results of the similar calculations with the neutron spectra from the LA150 library[10], shown in Fig. 2a, b as dash-dotted lines, are in obvious disagreement with the experiment.

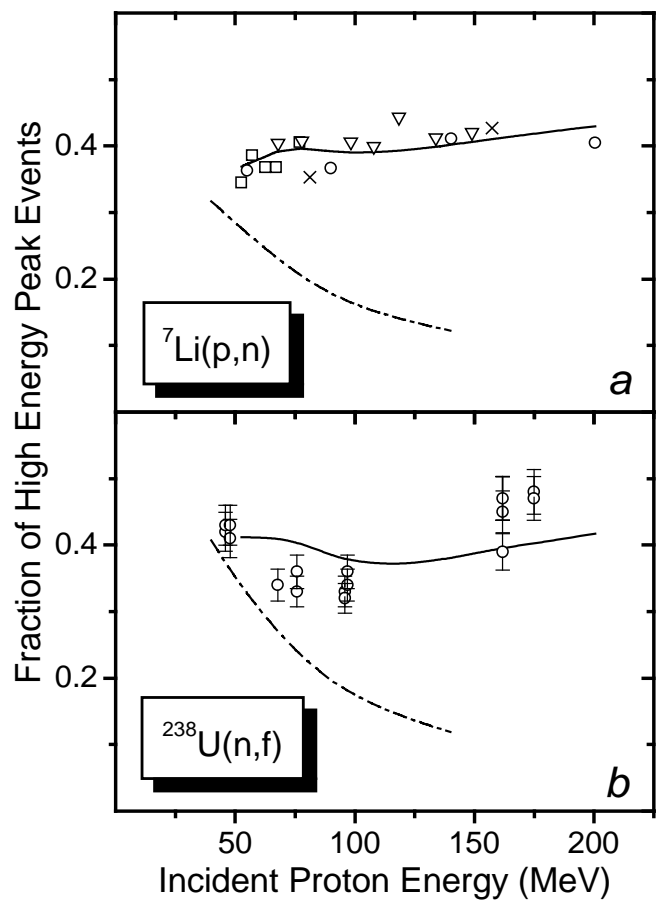

(a) Fraction of the high-energy peak in the neutron spectrum from the $^7$Li(p,n) reaction at 0°, versus incident proton energy. The solid line shows predictions of the systematics of the present work. The dash-dotted line corresponds to the spectra from the LA150 library[10]. The symbols represent the values deduced from the experimental data[2, 3, 15, 17], with the same designations as in Fig. 1.

(b) Fraction of fission events of $^{238}$U induced by high-energy peak neutrons from the $^7$Li(p,n) reaction at 0°, versus incident proton energy. The solid and dash-dotted lines represent the calculations that use, correspondingly, the systematics of the present work and the LA150 library data[10], folded with the standard $^{238}$U(n,f) cross section[12, 13]. The symbols show experimental data[21].

## V. Summary and future work

The systematics have been developed for the neutron spectra from the $^7$Li(p,n) reaction at 0° in the 50-200 MeV incident energy region. Reasonable description of the spectra is provided from the highest neutron energies down to ~10-20 MeV, which should be sufficient for most applications. The systematics should be used with caution at lower neutron energies because of the above mentioned effects that may influence the spectrum.

The systematics may be helpful for characterization of the existing $^7$Li(p,n) neutron beam facilities, as well as for the design of new ones. It is planned to use the systematics for further development of the $^7$Li-proton file in the LA150 nuclear data library. The next step in this direction is description of angular distribution of the secondary neutrons. Sparsely available experimental data in the continuum part of the spectrum indicate that Kalbach systematics[22] may be suitable for this task, with free parameters to be chosen so that to reproduce the spectra at 0° provided by the systematics of the present work.

**Acknowledgement**

We are thankful to M. Baba, R.C. Byrd, N. Nakao, H. Schuhmacher, W. Scobel, and T. Rönnqvist for providing their numerical data. Helpful discussions with H. Condé, J. Blomgren, R. Michel, and S. Neumann are acknowledged.